\documentstyle[epsfig,aps,prd,twocolumn]{revtex}

\def\mbar{{\overline m}}
\def\etal{{\it et al.}}
\def\ee{\hbox{$e^+ e^-$}}

\def\ie{{\it i.e.}}

\def\lsim{\mathrel {\vcenter {\baselineskip 0pt \kern 0pt
    \hbox{$<$} \kern 0pt \hbox{$\sim$} }}}
    \def\gsim{\mathrel {\vcenter {\baselineskip 0pt \kern 0pt
    \hbox{$>$} \kern 0pt \hbox{$\sim$} }}}

\begin{document}

\title{Temporary Acceleration of Electrons While Inside an Intense 
Electromagnetic Pulse}

\author{Kirk T.~McDonald}
\address{Joseph Henry Laboratories, Princeton University, Princeton, NJ 08544}

\author{Konstantine Shmakov}

\address{GlobalStar, Inc., San Jose, CA 95461}

\date{Nov.~20, 1997} 

\maketitle

\begin{abstract}
A free electron can temporarily gain a very significant amount of energy if it 
is overrun by an intense electromagnetic wave.   In principle, this process
would permit large enhancements in the center-of-mass energy of 
electron-electron, electron-positron and electron-photon interactions if these
take place in the presence of an intense laser beam.  Practical considerations
severely limit the utility of this concept for contemporary lasers incident on
relativistic electrons.  A more accessible laboratory phenomenon is
electron-positron production via an intense laser beam incident on a gas.
Intense electromagnetic pulses of astrophysical origin can lead to very
energetic photons via bremsstrahlung of temporarily accelerated electrons.

\noindent PACS numbers: 03.65.Sq, 12.15.-y  41.75.Fr, 52.40.Mj,  97.30.-b
\end{abstract}


The prospect of acceleration of charged particles by intense plane
electromagnetic waves has excited interest since the suggestion by Menzel
and Salisbury \cite{Menzel} that this mechanism might provide an explanation 
for the origin of cosmic rays.  However, it has generally been recognized that
if a wave overtakes a free electron, the latter gains energy from the wave
only so long as the electron is still in the wave, and reverts to its
initial energy once the wave has past \cite{Francia,Kibble,Sarachik,Kaw}.
There is some controversy as to the case of a ``short'' pulse of radiation,
for which modest net energy transfer between a wave and electron appears
possible \cite{Lai,Bucksbaum87,Jonsson,Malka,Wang}.  Acceleration via radiation
pressure is negligible \cite{Hagenbuch}.
It has been remarked that even in the case of a ``long'' pulse, 
some of the energy transferred from the wave to the 
electron can be extracted if the electron undergoes a scattering process while
still inside the wave \cite{Kibble,Kaw}.  This paper is an elaboration of that
idea.  We do not discuss here the observed phenomenon that an electron
ionized from an atom in a strong wave can emerge from the wave with
significant energy \cite{Bucksbaum90}.

We consider 
a plane electromagnetic wave (often called the background wave) with
dimensionless, invariant field strength 
\begin{equation}
\eta  = {e\sqrt{\langle{A_\mu A^\mu}\rangle} \over mc^2}
      = {e{\cal E}_{\rm rms} \over m\omega_0 c}
      = {e{\cal E}_{\rm rms} \lambdabar_0 \over m c^2}.
\label{eq2}
\end{equation}
Here the wave has laboratory frequency $\omega_0$,
reduced wavelength $\lambdabar_0$,
root-mean-square electric field ${\cal E}_{\rm rms}$,
and four-vector potential $A_\mu$; $e$ and $m$ are the charge and mass of the 
electron, and $c$ is the speed of light.

A practical realization of such a wave is a laser been.   Laser
beams with parameter $\eta$ close to one have been used in recent 
plasma-physics experiments \cite{Malka} and in high-energy-physics experiments 
\cite{Bula,Burke}.

When such a wave overtakes a free electron, the latter undergoes 
transverse oscillation (quiver motion), with relativistic
velocities for $\eta \gsim 1$ 
\cite{Francia,Kibble,Sarachik,Kaw,Landau,McMillan}.
The {\bf v} $\times$ {\bf B} force then couples the
transverse oscillation to a longitudinal drift in the direction of the wave.
In the nonrelativistic limit, this effect is often said to be due to the
``ponderomotive potential'' associated with the envelope of the electromagnetic
pulse \cite{Kibble}.  The resulting temporary
energy transfer to the longitudinal motion of the electron 
can in principle be arbitrarily large.

A semiclassical description of this process exists as well.
A quantum-mechanical electron inside a classical plane wave can be described
by the Volkov solutions to the Dirac equation \cite{Volkov,Lifshitz}.  Such
electrons are sometimes described as ``dressed'', and they
can be characterized by a quasimomentum
\begin{equation}
q = p + \epsilon k_0,
\label{eq4}
\end{equation}
where the invariant $\epsilon$ is given by
\begin{equation}
\epsilon = {m^2 \eta^2 \over 2 (p \cdot k_0)},
\label{eq5}
\end{equation}
with $(p \cdot k_0)$ being the 4-vector product of the 4-momenta $p$ of the
electron and $k_0$ of a photon of the background wave.
The factor $\epsilon$ need not be an integer, and can be thought of as an 
effective number of wave photons ``dragged'' along with the electron as a
result of a small difference between the large rates of absorption and
emission (back into the wave) of wave photons by the electron. 
(Strictly speaking, 
the wave used in the Volkov solution is classical and, hence, contains no
photons.)  As a result, the electron inside the wave has an effective mass, 
$\mbar$, that is greater than its free mass $m$ \cite{Kibble}:
\begin{equation}
\mbar^2 = q^2 = m^2(1 + \eta^2).
\label{eq6}
\end{equation}

From a classical view, the quasimomentum $q$ is the result of averaging over
the transverse oscillations (quiver motion) of the electron in the background
wave.  When discussing conservation of energy and momentum in the classical 
view, both
transverse and longitudinal motion of the electron must be considered; but
in a quantum analysis, quasimomentum is conserved and no mention is made
of the classical transverse oscillations.

Throughout this paper the background wave propagates in the $+z$ direction, and
the 4-momentum of a photon of this wave is written
\begin{equation}
k_0 = (\omega_0,0,0,\omega_0).
\label{eq3}
\end{equation}
From now on, we use units in which $c$ and $\hbar$ equal one.

We first consider a relativistic electron moving along the $+z$ axis with
4-momentum
\begin{equation}
p = (E,0,0,P) = \gamma m (1,0,0,\beta),
\label{eq1}
\end{equation}
where $E$ and $P$ are the energy and the momentum of the
electron prior to the arrival of the wave, 
$\beta \approx 1$ is the electron's velocity and  $\gamma = 
1/\sqrt{1 - \beta^2} \gg 1$.  Then
\begin{equation}
(p \cdot k_0) = \omega_0 (E - P) = {m^2 \omega_0 \over E + P},
\label{eq7}
\end{equation}
so
\begin{equation}
\epsilon = {\eta^2 (E + P) \over 2\omega_0} 
\approx {\gamma m \eta^2 \over \omega_0},
\label{eq8}
\end{equation}
where the approximation holds for a relativistic electron.
For a wave of optical frequencies (such as a laser), $\epsilon \gg 1$.  
The quasienergy, $q_0$, is then large:
\begin{equation}
q_0 = E (1 + \eta^2).
\label{eq9}
\end{equation}
The electron has been accelerated from energy $E$ outside the
wave to energy $E(1 + \eta^2)$ inside the wave.  Since $\eta$ can in principle
be large compared to 1, this acceleration can be very significant.

Can we take advantage of this acceleration in a high-energy-physics experiment?
The example of Compton scattering of an electron by one laser beam while in
a second laser beam has recently been reported elsewhere \cite{Hartemann}.
Here, we consider examples of possibly enhanced production of electroweak
gauge bosons in high-energy $ee$ and $e \gamma$  collisions in the presence
of an intense laser.


Suppose the electron $p$ collides head-on with a positron $p'$, all inside the 
background wave. The positron 4-momentum is then
\begin{equation}
p' = (E',0,0,-P'),
\label{eq10}
\end{equation}
where $E' \gg m$ in the relativistic case.  Then
\begin{equation}
(p' \cdot k_0) = \omega_0 (E' + P') \approx 2 E' \omega_0.
\label{eq11}
\end{equation}
The corresponding quasimomentum is
\begin{equation}
q' = p' + \epsilon' k_0,
\label{eq12}
\end{equation}
where
\begin{equation}
\epsilon' = {m^2 \eta^2 \over  2 (p' \cdot k_0)} 
            \approx {m^2 \eta^2 \over  2 E' \omega_0}.
\label{eq13}
\end{equation}
The factor $\epsilon'$ is not large in general, and the energy of a 
relativistic positron
(or electron) moving against an optical wave is almost unchanged.

However, the center-of-mass (cm) energy of the \ee\ system is increased when
the collision occurs inside the background wave.
The cm-energy squared is
\begin{equation}
s = (q + q')^2 
   \approx  4EE' (1 + \eta^2),
\label{eq14}
\end{equation}
which is enhanced by a factor $1 + \eta^2$ compared to the case of no
background wave.

For example, the $Z^0$ boson could be produced in \ee\ collisions with 
33- rather than 46.6-GeV beams, if the collision took place inside a background
wave of strength $\eta = 1$.

Of course, the background
wave Compton scatters off the positron beam at a high
rate if $\eta \gsim 1$, which results in substantial smearing of the energy of
that beam. In practice, the cm-energy enhancement by a background wave
would not be very useful in \ee\ or $ee$ collisions.

Note, however, that Compton scattering is insignificant when the background
wave and electron move in the same direction, unless the wave is
extraordinarily strong.  By an application of the Larmor 
formula in the (average) rest frame of the electron, we find that the
fraction of the electron's (laboratory) energy radiated in one cycle of its 
motion in the wave is of order $\alpha \eta^2 (\omega_0/E)$,
where
$\alpha$ is the fine-structure constant.


Suppose instead that the electron collides head-on with a high-energy photon 
of frequency
$\omega$ and 4-momentum
\begin{equation}
p' = k = (\omega,0,0,-\omega).
\label{eq15}
\end{equation}
Then eq.~(\ref{eq14}) holds on
substituting $\omega$ for $E'$; the cm-energy squared is
again enhanced by the factor $1 + \eta^2$.

The background wave can, of course, interact directly with the high-energy 
photon to 
produce \ee\ pairs, but if $4 \omega \omega_0 < m^2 (1 + \eta^2)$, the 
pair-production rate is much suppressed \cite{Burke}.  
Thus, there is a regime in which 
$e$ + photon collisions in a strong background wave are cleaner than \ee\
 or $ee$ collisions in the wave.

In practice, we could get the high-energy photon from Compton scattering of
the background wave off an electron beam.  One might not want to backscatter 
the wave
off a positron beam because of ``backgrounds'' from $e^+ e^- \to Z^0$.

A physics topic  of interest would be the reaction
\begin{equation}
k + e^- \to W^- + \nu,
\label{eq16}
\end{equation}
which proceeds via the triple-gauge-boson coupling $\gamma WW$, and whose
angular distribution is sensitive to the magnetic moment of the $W$ boson
\cite{Mikaelian,Ginzburg}.
The ``background'' process
\begin{equation}
k + e^- \to  Z^0 + e^-
\label{eq17}
\end{equation}
could be suppressed by suitable choice of polarization of the electron
and background wave.

For electron beams of 46.6 GeV as at the Stanford Linear Accelerator Center,
green laser light backscatters into
photons of energies up to about 30 GeV.  Thus if the laser had $\eta = 1$, the
cm energy would extend up to 106 GeV, well above the threshold for reactions 
(\ref{eq16}-\ref{eq17}).


However, the enhancement factors $1 + \eta^2$ in the
electron energy, eq.~(\ref{eq9}), and in the cm-energy squared,
eq.~(\ref{eq14}), of $ee$ or electron-photon
collisions are very much dependent on the idealization that the background
wave is highly collinear with the electron.  

We reconsider the preceding, but now suppose that the electron
makes angle $\theta \ll 1$ to the $z$ axis,  The 4-momentum of the
electron is 
\begin{equation}
p = (E, P \sin\theta,0,P \cos\theta),
\label{eq18}
\end{equation}
and
\begin{equation}
(p \cdot k_0) = E \omega_0 (1 - \beta \cos\theta)
 \approx {m \omega_0 \over 2 \gamma} (1 + \gamma^2 \theta^2).
\label{eq19}
\end{equation}
As a consequence, the (quasi)energy of the electron inside the wave is now
\begin{equation}
q_0 = p_0 + {m^2 \eta^2 \omega_0 \over 2 (p \cdot k_0)}
      \approx E \left( 1 + {\eta^2  \over 1 + \gamma^2 \theta^2} \right), 
\label{eq20}
\end{equation}
which reduces to eq.~(\ref{eq9}) as $\theta$ goes to zero.  
However, if $\theta > \eta/\gamma$, then
the electron is hardly accelerated by the background wave.

Electrons of present interest in high-energy physics typically have energies
in the range 1-1000 GeV, corresponding to $\gamma \approx 10^3$-$10^6$.  This
places very severe requirements on the alignment of the background wave with
the electron beam.  Indeed, the angular divergence of an electron beam is
often larger than $1/\gamma$, so that no alignment of the background wave
could impart large energy enhancements to the entire beam.

Furthermore, optical waves with $\eta \approx 1$ can only be obtained at
present in focused laser beams for which the characteristic angular spread is
$\Delta\theta \gsim 0.1$.  So even if the central angle of the beam could be
aligned to better than $1/\gamma$, only a very small fraction of the beam
power would lie within a cone of that angle.

We also note that for the quasimomentum $q$ to be meaningful,
the electron must have resided inside the strong background field for at least
one cycle.  A relativistic electron moves distance $2\gamma^2 (1 + \eta^2) 
\lambda_0$ while the background wave advances one wavelength relative
to the electron \cite{Chan}.  
However, the strong-field region of a focused laser 
is characterized by its Rayleigh range, which is typically a few hundred
wavelengths when $\eta \approx 1$.  Further, the transverse extent of the 
(classical) trajectory is of order $\gamma \eta \lambda_0$.
Hence, in present laser systems, the
strong-field region is not extensive enough that the energy transfer
(\ref{eq9}) could be realized for $\gamma \gsim 10$.



While physical consequences of the temporary acceleration of relativistic 
electrons inside 
an intense laser beam may be difficult to demonstrate, there is also
interest in the case where the electron is initially at rest, or nearly so,
such as electrons ionized from gas atoms by the passage of the background 
laser pulse \cite{Bucksbaum90}.

An interesting process is so-called trident production,
\begin{equation}
e + A \to e' + A' + e^+e^-,
\label{eq21}
\end{equation}
of an electron-positron pair in the interaction of an ionization electron
with a nucleus $A$ of a gas atom.
For a very heavy nucleus $A$, its final state $A'$ has a different momentum
but the same energy.  Then the initial electron must provide the energy to
create the $e^+e^-$ pair as well as that for the final electron.  The
least energy required is when all three final-state electrons and positrons
are at ``rest'' (\ie, they have zero net longitudinal momentum; they must always
have quiver motion when they are in the wave).
Thus, the minimum total quasienergy of the final-state electrons and positrons 
is $3\mbar$.

We conclude that the quasienergy
$q_0$ of the initial electron must be at least $3\mbar$ for reaction
(\ref{eq21}) to occur.

           
If the electron is at rest prior to the arrival of the background wave
its 4-momentum is 
\begin{equation}
p = (m,0,0,0).
\label{eq22}
\end{equation}
As the electron is overtaken by a wave of strength $\eta$ and
4-momentum given by (\ref{eq3}), it takes on quasimomentum
\begin{equation}
q = (m(1+\eta^2/2),0,0,m\eta^2/2) \equiv (\mbar\gamma,0,0,\mbar\gamma\beta_z).
\label{eq23}
\end{equation}
Thus, the net longitudinal velocity of the electron inside the
wave is $\beta_z = q_z/q_0 = (\eta^2/2)/(1 + \eta^2/2)$.  As expected, inside a 
very strong wave the electron can take on relativistic longitudinal motion.

We could have trident production while the electron is still in the wave
if the quasienergy $q_0 = m(1 + \eta^2/2)$ exceeds $3\mbar$.
For an electron initially at rest, this requires
$\eta \ge \sqrt{16+12\sqrt{2}} = 5.74$.

The trident process is still possible within a wave with $\eta < 5.74$ 
provided the electron has quasienergy $q_0 \ge 3\mbar$.  This might arise,
for example, because of acceleration of the electron by the plasma-wakefield 
effect \cite{Umstadter}.

It is conceivable that the electron creates the pair in a linearly
polarized wave at a phase when its (classical) kinetic energy is high, but the 
final electron and the pair all appear with a lower kinetic energy corresponding
to some other phase of the wave.  This can't happen if the interaction takes
place at a well-defined point, since the phase of the wave is a unique function 
space and time.  It might occur if the final particles ``tunnel'' to 
another space-time point before appearing, and the instantaneous kinetic
energy is lower at that point.

However, we will find shortly that such tunneling is not consistent with
energy conservation.  To be as definite as possible, we consider ordinary
energy along the classical trajectories, rather than quasimomentum.
The latter is taken into account in the sense that the electron and positron
are not created at rest, but with the transverse velocities appropriate to
phase of the background wave at the spacetime point at which the pair appears.
It is sufficient to consider only those trajectories with zero average 
momentum (\ie, zero quasi-3-momentum).

For circular polarization of the background wave,
 the electron trajectory is a circle
in the plane perpendicular to the $z$ axis, with radius $a/\omega_0$,
velocity $\beta = a$ and Lorentz factor
\begin{equation}
\gamma_{\rm circ} = {1 \over \sqrt{1 - a^2}} = \sqrt{1 + \eta^2},
\label{eq26}
\end{equation}
where parameter $a$ is given by
\begin{equation}
a^2 = {\eta^2 \over 1 + \eta^2}, \qquad 0 \leq a^2 \leq 1.
\label{eq25}
\end{equation}

For a background wave that is linearly polarized in the 
$x$-direction, the trajectory can be
parametrized as \cite{Sarachik,Landau}
\begin{equation}
x = \sqrt{2} {a \over \omega_0} \sin\delta, \qquad
z = {a^2 \over 4 \omega_0} \sin 2\delta,
\label{eq27}
\end{equation}
where $\delta = \omega_0\tau \sqrt{1 + \eta^2}
= \omega_0\tau /\sqrt{1 - a^2}$, 
and $\tau$ is the proper time.
Expression (\ref{eq27}) describes the well-known figure-8 trajectory.  Now
$dx/d\tau = (dx/dt)(dt/d\tau) = \gamma \beta_x$, 
so $\gamma^2 = 1 + \gamma^2\beta^2 = 1 + (dx/d\tau)^2 + (dz/d\tau)^2$.  
We find that
\begin{equation}
\gamma_{\rm lin} = {1 + {1 \over 2}[a^2 - (\omega_0 x)^2] \over \sqrt{1 - a^2}}.
\label{eq28}
\end{equation}
From expression (\ref{eq27}) for the $x$-trajectory we see that 
$0 \leq (\omega_0 x)^2 \leq 2a^2$, so
\begin{equation}
\gamma_{\rm min} 
= {1 + \eta^2/2 \over \sqrt{1 + \eta^2}}, \qquad \hbox{and}
\qquad \gamma_{\rm max} 
= {1 + 3\eta^2/2 \over \sqrt{1 + \eta^2}}. 
\label{eq29}
\end{equation}
These values surround the result that $\gamma_{\rm circ} 
= \sqrt{1 + \eta^2}$
always for circular polarization.  For small $\eta$, $\gamma_{\rm min} \approx
1 + \eta^4/8$, $\gamma_{\rm max} \approx 1 + \eta^2$, and $\gamma_{\rm circ}
\approx 1 + \eta^2/2$; for large $\eta$, $\gamma_{\rm min} \approx \eta/2$,
 $\gamma_{\rm max} \approx 3\eta/2$, and $\gamma_{\rm circ} \approx \eta$.

Suppose an electron interacts with a nucleus at the place where its
Lorentz factor is $\gamma_{\rm max}$ and
reappears along with an electron-positron pair at a location where
$\gamma_{\rm min}$ holds at that moment.  The nucleus absorbs the excess
momentum of the initial electron.  Conservation of (ordinary) energy requires 
that $\gamma_{\rm max} = 3\gamma_{\rm min}$.
But this is not satisfied for any value of $\eta$ according to (\ref{eq29}).
That is, the hypothetical tunneling process is not possible under any
circumstances.

In sum, even when in a background wave an electron can produce positrons off 
nuclei only if
the electron has sufficient longitudinal momentum that the corresponding
(quasi)energy is three times the (effective) electron mass.

We close by returning to the astrophysical context that began the historical 
debate on acceleration by intense electromagnetic waves.
Gunn and Ostriker \cite{Ostriker} have given an extensive discussion 
the possibility of electron acceleration in the rotating dipole field
of a millisecond pulsar,  where the field strength $\eta$ can be of order 
$10^{10}$.  Their argument
does not primarily address free electrons overtaken by a wave, but rather
electrons ``injected'' or ``dropped at rest' into the wave.
Neutron decay is a candidate process for injection.  In very strong fields
($\eta \gg 1)$ this decay takes place together with the absorption by the 
electron (and proton) of a very large number of wave photons, so that the 
electron is created with (quasi)energy $\approx m \eta^2/2$ (compare 
eq.~(\ref{eq23})) \cite{Nikishov}.  Because the fields of the pulsar fall off 
as $1/r$ where (coincidentally) $r_{\rm pulsar} \approx \lambda_0$, 
the wavelength of the rotating dipole radiation, the
field region is ``short'', and the electron may emerge with some
fraction of the large energy it had at the moment of its creation.

An example closer to the theme of the present paper would be an electron
that is overtaken by the 
intense electromagnetic pulse of a supernova (or other transient astrophysical
occurrence, perhaps including gamma-ray bursters), 
and thereby temporarily accelerated to energy $m \eta^2/2$.  
Such pulses could have significant fields at optical frequencies, where the
transverse scale, $\eta\lambda_0$, of the motion of accelerated electrons
is less than the Chandrasekhar radius for $\eta < 10^{10}$.
In general, the electron has low energy before and after the passage of the
pulse.  However, high-energy photons can arise via bremsstrahlung of the
electron when it interacts with a plasma nucleus while still in the pulse.
  In this view, the primary
astrophysical evidence of temporarily accelerated electrons would be 
high-energy photons which, of course, could transfer some of their energy to 
protons and other charged particles in subsequent interactions.

This work was supported in part by
DoE grants DE-FG02-91ER40671 and DE-FG05-91ER40627.



\begin{thebibliography}{[99]}


\bibitem{Menzel}
D.H.~Menzel and W.W.~Salisbury,
Nucleonics, {\bf 2}, No.~4, 67 (1948).

\bibitem{Francia}
G.~Toraldo di Francia,
Nuovo Cim.~{\bf 37}, 1553 (1965).

\bibitem{Kibble}
T.W.B.~Kibble,
Phys.~Rev.~{\bf 138}, B740, (1965); {\bf 150}, 1060 (1966);
Phys.~Rev.~Lett.~{\bf 16}, 1054 (1966);
{\sl Carg\`ese Lectures in Physics}, Vol.~2, ed. by M.~L\'evy (Gordon and
Breach, New York, 1968), p.~299.

\bibitem{Sarachik}
E.S.~Sarachik and G.T.~Schappert,
Phys.~Rev.~D {\bf 1}, 2738 (1970).

\bibitem{Kaw}
P.E.~Kaw and R.M.~Kulsrud,
Phys.~Fluids {\bf 16}, 321 (1973).

\bibitem{Lai}
H.M.~Lai,
Phys.~Fluids {\bf 23}, 2373 (1980).

\bibitem{Bucksbaum87}
P.H.~Bucksbaum \etal,
Phys.~Rev.~Lett.~{\bf 58}, 349 (1987).

\bibitem{Jonsson}
L.~J\"onsson,
J.~Opt.~Soc.~Am.~B{\bf 4}, 1422, (1987).

\bibitem{Malka}
G.~Malka \etal,
Phys.~Rev.~Lett.~{\bf 78}, 3314 (1997).

\bibitem{Wang}
J.X.~Wang \etal,
Phys.\ Rev. E {\bf 58}, 6575 (1998).

\bibitem{Hagenbuch}
K.~Hagenbuch,
Am.~J.~Phys.~{\bf 45}, 693 (1979).

\bibitem{Bucksbaum90}
P.H.~Bucksbaum,
in {\sl Atoms in Strong Fields}, 
C.~Nicholaides \etal\ eds.~(Plenum Press, 1990), p. 381.

\bibitem{Bula}
C.~Bula \etal,
Phys.~Rev.~Lett.~{\bf 76}, 3116 (1996).

\bibitem{Burke}
D.L.~Burke \etal,
Phys.~Rev.~Lett.~{\bf 79}, 1626 (1997).

\bibitem{Landau}
L.~Landau and E.M.~Lifshitz, 
{\sl The Classical Theory of Fields},
4th ed.~(Pergamon Press, Oxford, 1975), prob.~2, \S~47; p.~112 of the 
1941 Russian edition.

\bibitem{McMillan}
E.M.~McMillan,
Phys.~Rev.~{\bf 79}, 498 (1950).

\bibitem{Volkov} D.M.~Volkov,
Zeits.~f.~Phys.~{\bf 94}, 250 (1935).

\bibitem{Lifshitz} 
See also Secs.~40 and 101 of V.R.~Berestetskii, E.M.~Lifshitz and 
L.P.~Pitaevskii,
{\em Quantum Electrodynamics}, 2nd ed.~(Pergamon Press, Oxford, 1982).

\bibitem{Hartemann}
F.W.~Hartemann,
Phys.\ Plasmas {\bf 5}, 2037 (1998).

\bibitem{Mikaelian}
K.O.~Mikaelian,
Phys.~Rev.~D {\bf 17}, 750 (1978); {\bf 30}, 1115 (1984).

\bibitem{Ginzburg} 
I.F.~Ginzburg \etal,
Nucl.~Phys.~{\bf B228}, 285 (1983).

\bibitem{Chan}
Y.W.~Chan,
Phys.~Lett.~{\bf 35A}, 305 (1971).

\bibitem{Umstadter}
D.~Umstadter \etal,
Science, {\bf 273}, 472 (1996).
         
\bibitem{Ostriker}
J.P.~Ostriker and J.E.~Gunn,
Ap.~J.~{\bf 157}, 1395 (1969); {\bf 165}, 523 (1971).

\bibitem{Nikishov}
A.I.~Nikishov and V.I.~Ritus,
Sov.~Phys.~JETP {\bf 19}, 1191 (1964);
{\bf 58}, 14 (1983);
V.I.~Ritus,
{\it ibid.}, {\bf 29}, 532 (1969).


\end{thebibliography}
\end{document}